\documentclass{article}
\usepackage{epsfig,rotating}
\usepackage{subfig}
\usepackage{amsmath}
\usepackage{color}
\usepackage{verbatim}
\protect{\markright{Hierarchical Multiobjective Optimization}}

\begin{document}
%\linenumbers

\baselineskip 24pt
\title{A Hierachical Evolutionary Algorithm for Multiobjective
  Optimization in IMRT} 
\author{
Clay Holdsworth, Ph.D. \\
Minsun Kim, Ph.D. \\
Jay Liao, M.D. \\
Mark H Phillips, Ph.D.
\\
Department of Radiation Oncology, \\
University of Washington Medical Center, \\
Box 356043, Seattle, WA 98195, USA \\
}

\maketitle

\normalsize

\begin{abstract}

{\bf Purpose}:
Current inverse planning methods for IMRT are limited because they are
not designed to explore the trade-offs between the competing
objectives between the tumor and normal tissues.  Our 
goal was to develop an efficient multiobjective optimization algorithm
that was flexible enough to handle any form of objective function and
that resulted in a set of Pareto optimal plans.\\
{\bf Methods}: 
We developed a hierarchical evolutionary multiobjective algorithm 
designed to quickly generate a small diverse Pareto optimal set
of intensity modulated radiation therapy (IMRT) plans that meet all
clinical constraints and reflect the optimal trade-offs in any
radiation therapy plan.  
The top level of the hierarchical algorithm is a 
multiobjective evolutionary algorithm (MOEA).  The genes of the
individuals generated in the MOEA are the parameters that define the
penalty function minimized 
during an accelerated deterministic IMRT optimization that represents
the bottom level of the hierarchy.  The MOEA incorporates clinical
criteria to restrict the search space through protocol objectives and then uses Pareto 
optimality among the fitness objectives to select individuals.
The population size is not fixed, but a specialized niche effect, domination advantage, is
used which  controls the patient size and plan diversity.  The number 
of fitness objectives is kept to a minimum for greater selective
pressure, but the number of genes is expanded for flexibility that
allows a better approximation of the Pareto front. \\   
{\bf Results}:
Acceleration techniques implemented on both levels of the hierarchical
algorithm resulted in short, practical runtimes for multiobjective
optimizations.  The MOEA improvements were evaluated for example
prostate cases with one target and two OARs.  The 
modified MOEA dominated 11.3\% $\pm$ 0.7\% of plans using a standard
genetic algorithm package.  By implementing domination advantage and
protocol objectives, small and diverse populations of clinically
acceptable plans that were only dominated 0.2\% $\pm$ 0.05\% by the
Pareto front could be generated in a fraction of an hour.   \\
{\bf Conclusions}:
Our MOEA produces a diverse Pareto optimal set of plans that meet all
dosimetric protocol criteria in a feasible amount
of time.  Our final goal is to improve practical aspects of the algorithm and
integrate it with a decision analysis or human interface for selection
of the IMRT plan with the best possible balance of successful
treatment of the target with low OAR dose and low risk of complication
for any specific patient situation.  

\end{abstract}


\section{Introduction}
\label{introduction}

\vspace{-12pt}
Intensity modulated radiation
therapy (IMRT) uses inverse planning optimization to determine complex
radiation beam configurations that will deliver dose distributions to
the patient that are better than could be achieved using 3D
conformal radiation therapy.  The problem is inherently multiobjective
with competing clinical goals for a high, uniform target dose  and low
dose to the organs at risk (OAR).
The IMRT inverse planning algorithms commonly optimize a single
objective function (also known as "penalty function") that 
is a linear combination of many separate organ and
tumor-specific objective (penalty) functions.  The solution of this
problem is one point  in the space of feasible treatment plans.  To
find another point in this space,  the user can vary any one of the
parameters that make up the penalty function.  Current clinical algorithms do not
provide methods for efficiently searching this space of feasible
plans.  The search is complicated by the fact that in comparing plans,
one plan may have achieved superior values for some objectives but
inferior values for other objectives. 

A multiobjective optimization (MOO) algorithm is one method for searching
through the space of feasible plans.  Such an algorithm varies the
parameters that define an objective function and calculates a new plan
for each new function.  For each plan created, the algorithm
calculates objective values that characterize its quality.  If all of
the objective values of one plan are better or the same as the values
of a second plan, the first plan is said to dominate the second.  A
plan is said to be Pareto optimal if there are no plans that dominate
it.  Note that any Pareto optimal plan 
has some superior objective values compared to other Pareto optimal
plans, but inferior values for others.  MOO algorithms must generate a
large number of plans that each require an individual IMRT
optimization.  The resulting computation time could make this
algorithm impractical depending on the efficiency of IMRT optimization
method used and the number of plans required.  The final product of a
MOO algorithm is a set of Pareto optimal plans, known as the {\it
  Pareto front}, so ultimately, there must be a method to select the
individual plan to be used.  

The true Pareto front of this multiobjective problem is a
multidimensional surface that contains an infinite number 
of points, and hence there is the potential for the production of
Pareto populations that are too large for practical purposes.  Methods
for handling this issue, called {\it niche
functions}, come in many forms, but are all designed to limit the number
of individuals that can survive in a population while maintaining
diversity.  For stochastic multiobjective algorithms to be practical,
the use of an appropriate niche function may be quite important. 

The nature of the separate objective functions used in the MOO plays an
important role in the optimization.  Those functions that are convex
may be optimized using deterministic algorithms.  Stochastic
optimization methods are needed for optimizing those functions that
are not convex to avoid becoming trapped 
in local minima.  In selecting a multiobjective optimization approach,
the user may have to make a decision between using a deterministic
method that is fast but limits the type of objectives and a
stochastic method that is very flexible but slower.  If different
objective functions are used to evalute the quality of dose
distributions than are used in the single penalty
function that optimizes beamlet intensities, it is possible that some
parts of the optimization can be done deterministically while
other parts may use stochastic methods. 

One type of stochastic optimization algorithm is an evolutionary
algorithm which uses concepts based on genetics and natural selection
to create a population of new solutions and to select the fittest
among them.  An "individual" is described by a set of genes.  New
individuals are created through recombination of genes from two
"parent" individuals selected from the existing population.  Random 
mutations can occur which help prevent the population from becoming
trapped in local minima.  Natural selection of superior individuals
followed by reproduction using their genes leads the algorithm to find
the Pareto optimal solutions.  In practice, further heuristic methods
are often needed to speed up this process.  

This paper describes the development of a multiobjective optimization
algorithm that combines an evolutionary method to search feasible plan
space with a deterministic method using a single objective penalty
function that quickly finds the optimal  
beamlet intensities for that function.  While the penalty function
must be convex, the hierarchical method allows us to use 
any fitness objective (convex or non-convex) to evalute plan quality
without losing too much in 
speed.  Utilization of a MOO algorithm requires the selection of one
plan from the Pareto optimal set to be used for IMRT treatment of the
patient.  There are many methods for doing so.  We have selected an
{\it  a posteriori} method of multiobjective optimization where plan selection is done after the Pareto front is found by either an influence diagram
\cite{Meyer04,Smith09} or inspection by a radiation oncologist.  With
both of these approaches, it is best to restrict the set of plans to a
relatively small number, e.g. 10-20, and this is reflected in our MOO
algorithm.   

In the methods section,
the hierarchical algorithm is described with the emphasis on the evolutionary
element and how it interacts with the deterministic step.  In the results,
performance of the algorithm is characterized using two clinical
prostate cancer cases.  

\vspace{-12pt}
\section{Methods and Materials}

\vspace{-12pt}
A multiobjective evolutionary algorithm (MOEA) has been developed for
the top level of a hierarchical IMRT optimization method.  The lower level
is a deterministic algorithm that incorporates a constrained quadratic
optimization algorithm (BOXCQP).  
We have used a hierarchical process because there are two sets of
variables we would like to optimize: (1) beamlet intensities (deterministic optimization), and (2)
the objective parameters such as maximum dose, dose
homegeneity, etc (stochastic optimization).  
At the stochastic level of the algorithm, an individual is created and
each individual is characterized by a set of genes that determine the 
relative weighting between OARs and targets and a 
set of genes that describe the OAR penalty functions.
The deterministic algorithm incorporates these genes into 
a quadratic objective function which is then solved for the beamlet
intensities.  The MOEA uses a separate set of objectives to determine
the quality of each individual treatment plan.  As mentioned
above, considerations of speed have led us to incorporate several
heuristic stages in our algorithm which results in a faster
convergence to a clinically-acceptable set of Pareto optimal plans. 

The following sections provide more detail of the algorithm as well as
descriptions of the methods used to characterize the performance of the
MOEA.  

\vspace{-12pt}
\subsection{Multiobjective Evolutionary Algorithm}

\vspace{-12pt}
An evolutionary algorithm was chosen as the stochastic
component of the 
multiobjective method because of its effectiveness in searching large regions of
solution space while still incorporating techniques for focusing the
search on regions more likely to produce better plans  
\cite{Ezzell96,Wu01,Coello02,Osyczka02,Schreibmann04a}.  
A number of characteristics of the IMRT optimization problem guided 
the development of our MOEA.  The critical factors were (1) the
very large search space, (2) the relatively large number of
objectives, (3) the need for solutions to conform to 
"clinically-acceptable" criteria, and (4) the time constraints for
clinical use.  In short, our modified MOEA was designed for faster
and more effective convergence to the Pareto front inside the 
clinically acceptable space.  The elements of the
MOEA and highlights of the impact of these critical factors are described below.

\vspace{-12pt}
\subsubsection{Enhancements to the MOEA}

\vspace{-12pt}
Our MOEA uses no fixed number of individuals per
generation, no 
summary single fitness function, and no tournament style selection
method (as are common in traditional genetic  algorithms).  Instead,
individuals 
are created one at a time and are eliminated if they are Pareto
dominated by any existing individual or if they violate any clinical
constraints.  Conversely, if the new individual dominates any
of the current population, those dominated individuals will
be eliminated.  If the individual survives, its genes become
candidates for recombination in the next individual.  In addition to
these algorithmic modifications, various parameters of the
multiobjective method, such as mutation rate, have been optimized for
performance. 

We also implemented a specialized niche function,
called {\it domination advantage}.  When calculating whether plan A
dominates plan B, each objective value of plan A only needs to be
within an $\epsilon$ of the corresponding plan B objective in order to
dominate it.  This domination advantage, $\epsilon$, can be thought of
as the difference in objective values below which the clinical
significance cannot be determined.  Thus, it provided a means to avoid
clustering of solutions so that the Pareto optimal set spans
the entire clinically relevant space.  It was also used to limit the
total number of plans. The $\epsilon$ used in
domination advantage was not a constant but was set equal to $c \cdot
(n_{current} - n_{goal})$.  The constant $c$ was adjusted so that the
final number of plans was approximately the desired number,
$n_{current}~\sim~n_{goal}$.   

If two individuals were being compared to test for domination and were
very close together, i.e. with all objective values within $\epsilon$ of
each other, both individuals would 'dominate' the other using
domination advantage.  In this rare case, a single fitness function was
used that was a weighted sum of fitness objectives using importance
weighting predetermined by the user.  The individual with the worse
single fitness function would be eliminated and the other would
survive.  Note that the result is different from might occur in
traditional evolutionary algorithms where two individuals from
different regions of the objective space would be forced into such a
comparison.  In our algorithm, these two individuals are so close
together, the assumption is that little information of clinical value
would be lost by the undesirable, but necessary, use of a single
fitness function elimination. 

We used two sets of objectives in the stochastic  
part of the algorithm, which we call {\it fitness objectives} and {\it 
  protocol objectives}.  Fitness objectives are the dosimetric
functions that best represent the clinical goals and are used to
determined plan quality.  Protocol objectives are functions 
that delineate certain dosimetric cutpoints that need to be met,
usually because of clinical protocol restrictions.  Protocol objectives will
restrict the search space such that the final set of plans are
guaranteed to meet all clinical criteria.  As will be described in
more detail below, these two sets of objective functions are distinct
and play different roles during the optimization.

\vspace{-12pt}
\subsubsection{Fitness and Protocol Objectives}

\vspace{-12pt}
In multiobjective optimization, smaller numbers of fitness objectives
tend to result in greater selective pressure and better performance.
Thus, it is desirable to have only one fitness objective for each
structure.  We chose effective uniform dose (EUD) as an appropriate
objective for OARs; other functions such as complication probability
functions could also be used.  Unlike for OARs, we did not find target EUD
to be an effective objective, because a uniform target at
prescription dose could be deemed equivalent to a dose distribution
with high variance and a higher mean dose.  Typically, target dose
distributions must meet minimum dose requirements as well as limits on
inhomogeneity.  We removed the minimum dose
objective by prioritizing it and scaling the dose distributions to meet all
minimum dose requirements, thereby leaving maximum dose (dose
homogeneity) as the sole target objective.  

The two sets of objective functions that were used are: 

\begin{itemize}
\item Fitness objectives \\
\begin{equation}\label{eq:fitness}
\begin{split}
\vec{f}_{fit}(\vec{x}) & = (f_{fit,1}(\vec{x}),\cdots, 
 f_{fit,m}(\vec{x}), f_{fit,m+1}(\vec{x}),\mathellipsis, f_{fit,m+n}(\vec{x})), \\ 
\text{where}~ m & = \#~\text{of target fitness objectives} \\
   n & = \#~\text{of OAR fitness objectives} \\
f_{fit,i}(\vec{x})& = d^{max}_i \text{ for} ~i \in
\{1,\mathellipsis,m\} \text{, i.e. for targets}\\
 f_{fit,i}(\vec{x})& =~EUD(\vec{d})  ~ = ~
(1/nvox_i \sum\limits_{j=1}^{nvox_i}
d_j^{a_i})^{1/a_{i}} ~ \text{for}~ i~ \in~ \{m+1, \mathellipsis, m+n \} \text{, i.e. for OARs}. \\
\end{split}
\end{equation}
where $d_j$ is the dose to $j^{th}$ OAR voxel,
$nvox_i$ is the number of voxels in the $i^{th}$ OAR, 
 $d^{max}_i$ is the maximum dose to any voxel within the {\it i}th
  target, 
and $a_i$ is a constant ($>$ 1) that characterizes the radiation response of each of the $n$
OARs \cite{Niemierko97,Niemierko99,Luxton08}.

  \item Protocol objectives \\
\begin{equation}\label{eq:protocol}
\begin{split}
\vec{f}_{prot}(\vec{x}) & = (f_{prot,1}(\vec{x}),\cdots, 
 f_{prot,p}(\vec{x}), f_{prot,p+1}(\vec{x}),\mathellipsis, f_{prot,p+q}(\vec{x})), \\ 
\text{where}~ p & = \#~\text{of target protocol objectives} \\
   q & = \#~\text{of OAR protocol objectives} \\
  f_{prot,k} &~=~\begin{cases} d^{max}_k  & \text{if}~ d^{max}_k >
    D_{prot,k} \\ 0 & \text{otherwise} \end{cases} ~\text{  for all clinical maximum dose contraints} \\
  f_{prot,k} &~=~\begin{cases} nvox_{violate,k}  
    & \text{if}~ \frac{nvox_{violate,k}}{nvox_k} > V_{OAR,k}
    \\ 0 & \text{otherwise} \end{cases}
 ~\text{  for all clinical dose volume constraints}  \\
\end{split}
\end{equation}
where $D_{prot,k}$ is the maximum dose allowed allow to the structure associated with the $k^{th}$ protocol objective,
$nvox_{violate,k}$ is the number of voxels in 
the OAR associated with the $k^{th}$ protocol objective that receive doses that exceed the allowed dose of a
dose volume constraint (DVC), $nvox_k$ is the total number of voxels
in the OAR associated with the $k^{th}$ protocol objective, and $V_{OAR,k}$ is the maximum fraction of the OAR associated with the $k^{th}$ protocol objective
allowed over the dose for a DVC.   Minimum target dose constraints are
achieved by scaling the dose distribution if necessary.
\end{itemize}

\vspace{-12pt}
\subsubsection{Stages of the MOEA}
\label{stages}

\vspace{-12pt}
There are three stages of the MOEA that correspond to: (1) a broad search
of the feasible plan space, (2) a focusing of the search to a region
of space that meets protocol requirements and (3) a
search of the feasible space within the protocol constraints.
Throughout the last two stages, the domination advantage procedure is
used to both spread the plans out (avoiding clustering near local
minima) and to limit the number of plans.  As the optimization
proceeds, the current set of solutions approaches the true Pareto
front defined by the fitness objectives.

In stage 1, the individuals were
created by random selection of genes.  The genes were used to construct
the quadratic objectives of the deterministic part of the algorithm
(see Sec.~\ref{deterministic}).  Then the fitness and protocol objective
values were calculated, and the individual was added to or rejected from the
current population, whose size was $n_{current}$.  When $n_{current}$
equaled the pre-defined number of desired plans determined by the
decision procedure, $n_{goal}$, stage 1 ended. 

In stage 2, new individuals were created by 
recombination of current genes and mutation.  During this stage, the
direction of mutations were guided depending on whether or not the
parent plan met the protocol objectives in order to more quickly
search relevant space.  Selection proceeded as in
stage 1 with the addition of the domination advantage procedure.  In
this stage, the population was allowed to grow larger than $n_{goal}$.
Stage 2 ended 
when there were $n_{goal}$ individuals who met all of the clinical
criteria (protocol objective values = 0), at which time, all
individuals who did not meet these objectives were eliminated.  

Stage 3 proceeded similarly as stage 2 except that the 
protocol objectives were converted to constraints, mutation direction
was not guided, and Pareto optimality was
determined only by the fitness objectives.  This stage ran for a
pre-determined number of iterations.  

One reason to separate the objectives into two sets is to reduce the
dimensionality of the final solution space.  As the number of
fitness objectives considered in Pareto domination increases, the
probability of domination is reduced by a factor on the order of $\sim
.5^{n_{objectives}}$, and hence selective pressure was reduced.  By
converting protocol objectives 
into constraints, the solution space was restricted to a smaller
number of dimensions equal to the smaller number of fitness objectives.
 
A multiobjective optimization algorithm needs to be paired with a
decision making component.  This will have an impact on the number and
variation of the solution set.  In our case, the decision maker is
either a human or an influence diagram, neither of which can handle
large numbers of solutions or very nearly identical solutions.  Hence,
our algorithm was designed to produce on the order of $n_{goal}$
plans, i.e. 10-15 plans for $n_{goal}$ = 10.  The algorithm can be
adjusted to produce any size solution set. 

\vspace{-12pt}
\subsection{Deterministic Algorithm}
\label{deterministic}

\vspace{-12pt}
For the beamlet intensity optimization, we employed a
modification of the Bound Constrained Convex Quadratic Problem
(BOXCQP) algorithm \cite{Breedveld06,Voglis04,Phillips08}.  
This efficient method was implemented to minimumize a quadratic
objective function with the bound constraint that all beamlet
intensities are non-negative and less then some maximum allowed
beamlet intensity, $x_{max}$.  Non-zero penalty factors were assigned to
OAR voxels that had exceeded the OAR dose parameter in each
iteration.  A short description is provided.

To compute optimal beamlet intensities, BOXCQP
minimized the matrix form of the penalty function defined by each
individual's genes summed with a smoothing function as described below: \\ 
\begin{equation}\label{eq:deterministic}
\begin{split}
\mbox{minimize } ~f_{det}(\vec{x}) & \medspace =
(\vec{d}-\vec{D}_{det})^{T} P (\vec{d} -
\vec{D}_{det})|_{\vec{d}=B \cdot \vec{x}}  \\     
 & \qquad +~\kappa (S\vec{x})^{T}(S\vec{x})\\ 
 & \mbox{ subject to } 0 \leq x_b
\leq x_{max} \\
\end{split}
\end{equation}

where $f_{det}$ is the objective function for the determinstic
  component of the algorithm, $\vec{x} $ is the vector of the beamlet
  intensities, $B$ is the precalculated matrix storing the dose to 
each voxel from each beamlet, $P$ is a diagonal matrix of all
importance weighted penalty factors for each voxel, 
$\kappa$ is the smoothing constant, S is the smoothing penalty
matrix, and $x_b$ represents any element in vector $\vec{x}$ 

A smoothing term was introduced because BOXCQP depends
on solving a well-posed problem \cite{Breedveld06,Voglis04}.  It is
also desirable to find the deliverable solution for a 
beamlet weight vector $\vec{x}$ that is not excessively modulated due to
delivery limitations, i.e. MLC movement.  The smoothing term
was based on the fact that the smaller the second derivative of a fluence
map, the easier it is to deliver in practice.  Therefore, $S$ was the
second derivative matrix by finite difference method, and $\kappa$ was
a smoothing coefficient that was found empirically. 

The algorithm was implemented using
Matlab\footnote{The MathWorks, Natick, MA}.  
Since this code was executed many times during the course of the
optimization, considerable effort was taken to speed execution.  

\vspace{-12pt}
\subsection{Treatment planning platform}
\label{sec:Prism}

\vspace{-12pt}
A treatment plan consists of a set of beams, which describe the
geometry of radiation delivery.  Each radiation beam was divided
into a 2D array of beamlets.  Each beamlet represents a .6 cm by 1 cm 
area of the beam.  A beamlet length of 1 cm was selected because of
the size of the collimator leaf, and a width of 0.6 cm was used as a
compromise between speed and resolution.  Beamlet intensities were the
variables optimized in the deterministic IMRT optimization algorithm.   

The patient was considered as a collection of 3D
voxels that are segmented into a set of targets and critical
normal tissues (OARs).  A model of radiation
transport mapped each beamlet intensity to the dose it deposited at each
voxel; this was stored in a beamlet-to-voxel dose contribution matrix,
 which was calculated once before the optimization.  Typically
the number of beamlets ranged from 1000-5000 while the number of voxels
ranged from 10$^4$ to 10$^5$.  The entire system was modeled using
the Prism treatment planning system \cite{Kalet96,Kalet97,Kalet97a}
that was augmented with a pencil beam dose calculation algorithm
\cite{Phillips99}.  The optimization algorithms interacted with Prism
via file input/output. 

\vspace{-12pt}
\subsection{Experiments}

\vspace{-12pt}
As the focus of this work is to introduce this multiobjective method,
results are shown for only two prostate cancer example cases.  
Case 1, used in all experiments, was selected to be challenging and
had a PTV that was $\sim$8.5 cm across at the widest point and had
39\% overlap with the bladder and 32\% overlap with the rectum.  Case
2, used only in sections 3.5 and 3.6, had a PTV that was $\sim$7cm in
diameter with only 9\% overlap with the bladder and 17\% overlap with
the rectum.  All plans used seven equally spaced 6 MV photon beams.  

Three fitness objectives were used in all of the optimizations: bladder
(or bladder wall) EUD, rectal wall EUD, and target maximum dose.  OAR
voxels inside the PTV were excluded from objective calculation.  All 
dose distributions were scaled to meet minimum 
target dose requirements, thereby making the  target maximum dose a
measure of target variance.  The protocol objectives represented a
maximum allowed target dose and OAR DVH constraints.   A comparison
was also made for two example prostate patients between the MOEA and a
plan generated using a commercial objective deterministic IMRT
optimization algorithm with the current clinical protocol.  Of all
possible MOEA plans, we chose the one plan from the MOEA output that most
closely matched the clinical plan's target dose distribution. 

It should be noted that stochastic algorithms do not guarantee 
convergence to the Pareto optimal set.  The degree to which this
algorithm approximates that set is the subject of several of the
result sections described below.  

The following experiments were performed to evaluate individual
aspects of the algorithm:   
\vspace{-12pt}
\begin{itemize}
  \item demonstration of the evolution of a population of IMRT plans
    towards the Pareto front
    through the three stages of a multiobjective optimization (Section 3.1);
 \item a mulitobjective comparison of performance between a
   standard genetic algorithm package and the modified evolutionary 
    algorithm (Section 3.2); 
  \item demonstration of the role of domination advantage in
    controlling population size, maintaining diversity, and improving
    performance (Section 3.3);  
  \item evaluation of the performance of the MOEA with and without
    protocol objectives (Section 3.4);  
  \item a DVH comparison between plans produced by the MOEA and plans
    produced using our current clinical protocol and a commercial planning
    system (Section 3.5); 
  \item assessment of the ability of the MOEA
    to approximate the true Pareto front (Section 3.6). 
\end{itemize}

A common method for comparing two different algorithms is
to select the 'best' plan from each population and do DVH
comparisons; however, such a method is not a multiobjective metric that can
compare algorithms that produce populations of plans.  It 
focuses on only a small subset of the results, does not lead itself to
quantitative or statistical analysis, and, given the multiobjective
nature of the problem, selection of the plan used 
for comparison is subjective.  As previously discussed, in
multiobjective optimization a plan can only be deemed superior to
another if it is better in all of its objectives, which is the
definition of Pareto domination.  In our evaluation of different
multiobjective algorithms, we used a metric
based on Pareto dominance that we term "domination comparison", $C_D$.
This is the percentage of
all possible comparisons between two groups of plans in which the plan
from the first population dominates the plan from the second
population.  In mathematical form, the domination comparison metric
for a population resulting from one algorithm, $U$, 
in comparison with a population of plans obtained with another algorithm,
$V$, is defined as follows:  \\ 
\begin{equation}
\begin{split}
\text{Let the set of plans obtained by formulation}
~U~ & =~\vec{u} \\
\text{and those obtained by formulation} ~V~ & =
~\vec{v} \\
\text{and}~C_D~=~\frac{\sum_i \sum_j f(u_i,v_j)}{|U|*|V|} \\
\text{ where}~f(u_i,v_j) = 1 \text{	if } ~u_i \text{~dominates~} v_j \\
\text{and } f(u_i,v_j) = 0 ~\text{otherwise.}~ \\
\end{split}
\end{equation}

Domination comparison is a purely multiobjective metric that gives
some indication of which of two populations is closer to the Pareto
front; however, it does not give a measure of the diversity of the
population or coverage of the Pareto front.  For this reason we
included 3-D Pareto plots for graphical display of the final
populations for the two prostate examples.  For greater ease, we
made all of these plots uniform, with rectum objective on the x-axis,
bladder objective on the y-axis, and target objective in color scale. 

\vspace{-12pt}
\section{Results}
\label{results}

\vspace{-12pt}
\subsection{Stages of the MOEA}
\label{sec:phases}

\vspace{-12pt}
A population of plans for case 1 was tracked 
through the three stages of the evolutionary algorithm
(Figures 1a, 1b, and 1c).  Three fitness objectives were used: rectal wall EUD,
bladder EUD, and maximum target dose (MT).  Only OAR voxels outside
the target were used in the calculation of the fitness objectives.   

Often Pareto optimality is demonstrated using only two 
objectives which makes determining the Pareto front visually easy.
Given that we used three fitness objectives, we used color to indicate the
value of the third axis. For
uniformity, rectal wall EUD was plotted on  the X-axis, bladder
EUD on the Y-axs, and the value of the maximum target dose (MT) was
represented in color scale.  The plots illustrate the difficulty in 
visualizing even a 3-dimensional fitness objective space, which is
small compared the the dimensions of most IMRT planning.

Dose distributions were scaled to meet all minimum target dose
requirements to eliminate the need for a minimum target dose fitness
objective.  The maximum target dose represents clinically undesirable
hot spots and is a measure of target variance. Throughout the
presentations of these results, lower fitness objective values correspond to 
more optimal plans.  We used $n_{goal}$ = 10 throughout this work.  

The population after the first 10 iterations obtained from
completely random genes have relatively poor objective values. 
Stage 2 is complete after 53 iterations, and objectives values have
improved.  The ten clinically acceptable plans are shown
with a black box outline.  All other plans were removed before
stage 3 began.  For this example, the greatest amount of
improvement is seen in a small number of iterations during stage 2
as the initial population was not close to Pareto optimal.  
Stage 3 was completed after a total of 200 iterations.  Further improvement is
observed in the objective space during stage 3. 

\vspace{-12pt}
\subsection{Performance of the Hierarchical Evolutionary Algorithm}
\label{sec:performance}

\vspace{-12pt}
The modified evolutionary algorithm was compared to the Matlab standard genetic
algorithm package that used a single stage, a set population size, and
used a tournament-style selection method with a single fitness
function.  Both  
algorithms used the previously described deterministic algorithm.  To
isolate differences in the algorithms, domination advantage and
protocol objectives were not used in either
multiobjective optimization.  Important differences included the
selection method, allowed population size, and mutation rates.  Bladder EUD, rectal wall EUD and maximum target dose (MT) were
used as the fitness objectives.  

Twenty-five multiobjective optimizations that each produced a set of
11 to 15 plans were 
performed for both situations to establish statistical
significance, and each multiobjective optimization executed a total of 200
BOXCQP iterations in $\sim$10 minutes.  Objective values (OAR EUD and
target maximum dose) for individual plans resulting from two
optimizations are shown in Figure 2. 

Populations resulting from our MOEA dominated
populations resulting from the standard genetic algorithm $D_C$ = 11.3\%
$\pm$ 0.7\% of possible comparisons versus $D_C$ = 0.04\% $\pm$
0.02\% in reverse.  Results indicate that the new evolutionary
algorithm was able to produce a population of plans with lower OAR EUD
values and lower target variance in a short amount of time. 

\vspace{-12pt}
\subsection{Domination Advantage}
\label{sec:domadv}

\vspace{-12pt}
Our MOEA does not have a set population size or limit and
does not use a single-objective fitness function for selection.
Resulting populations could potentially be very large even with a
relatively small number of objectives.  To keep the population size in
the desired range, to increase 
selective pressure, and to promote diversity in the population, a
specialized niche effect, domination advantage, was used.  With
domination advantage, once the
population was greater than $n_{goal}$, potentially dominating individuals
were given an advantage when testing for domination.  One or more
objectives of a dominating individual can be inferior as long as it was
within $\epsilon$ of the objective of the individual being tested.
The size of $\epsilon$ increased linearly with the difference between the
$n_{current}$ and $n_{goal}$.  A comparison 
of two multiobjective optimizations of a prostate case (with
and without domination advantage) was performed 
 to demonstrate the effect on a single
population (Figure 3).  Twenty-five multiobjective
optimizations were performed with and without domination advantage to
establish statistical significance.

Average population size without domination advantage was 68.7 
$\pm$ 2.1 versus 13.2 $\pm$ 0.2 with domination
advantage. The population generated using the niche effect dominated
an average of $D_C$ = 1.73\% $\pm$ 0.35\% of the population generated
without it and $D_C$ = 0.17\% $\pm$ 0.04\% of dominations occurred in
reverse. In addition to creating a smaller, diverse population,
improvement in the resulting population was observed when domination
advantage was used.  When domination advantage was used, there were fewer
individuals for a more managable final population size, and 
individuals were closer to the true Pareto front possibly due to the
increase in selective pressure from the niche effect.  Domination
advantage did not allow searching in regions that give only a small 
improvement in one objective at great cost to other objectives.  This
speeds up the optimization; however, it also means that
some of the Pareto front may not be mapped.

\vspace{-12pt}
\subsection{Effect of Protocol Objectives}
\label{sec:protocolobj}

\vspace{-12pt}
Protocol objectives were used to focus the search to a region of
objective space that met pre-determined clinical dosimetric requirements.
Performance of multiobjective optimizations were
evaluted for an example prostate case with and without 
protocol objectives that reflected constraints from a clinical protocol.  The
effectiveness of the preference space guidance algorithm is shown in
Figure 4.  

In terms of Pareto domination the protocol objectives did not improve or degrade
results; however, without protocol objectives, the search focused on
too wide a search space.  Of all individual plans generated without
using protocol objectives, only 4.4\% $\pm$ 0.8\% met the clinical
criteria, and only 68\% of the full multiobjective optimizations had
at least one plan that met all clinical requirements.  In other words,
32\% of the optimization procedures did not yield a usable plan when
the protocol constraints were considered, whereas all of the plans
generated using the protocol objectives met all of the clinical criteria.

\vspace{-12pt}
\subsection{Two Clinical DVH Comparisons}
\label{sec:DVH}

\vspace{-12pt}
The MOEA was compared to our clinical planning methods for the two
prostate cases.  IMRT plans were optimized
using a 200 iteration MOEA optimization and using our standard
clinical inverse planning procedure\footnote{Pinnacle$^3$, Philips
  Medical Systems}.  The MOEA plan whose target dose distribution most
closely matched the plan using the clinic's deterministic algorithm
was chosen from each multiobjective population for each case.
A DVH comparison was made between these plans and the
corresponding clinical plans.  None of the example patient studies
included femur and unspecified tissue objectives, and the MOEA 
plans were not translated into deliverable beams.  This comparison is only a
demonstration of the potential of this algorithm and not designed to
be a comparison of plans ready to be used in the clinic. 

The DVH comparisons for the target, rectal wall, and bladder for
case 1 are shown in figures 5a-5c. 
The minimum target dose requirements were matched for these plans;
however, the uniformity of the target dose distribtuion and lower
maximum target dose were better for the MOEA plan.  As this patient
was more difficult especially with regards to the proximity of the
entire bladder with the target, improvement of the MOEA OAR
distributions over the clinical plan was modest. 32\% of the volume of
the rectum was inside the target, and some of this volume received
higher dose in the MOEA plan.  Once outside the PTV, the dose to the rectum dropped off more quickly for the MOEA than for the plan
produced using the current clinical protocol. 

The DVH comparisons for the target, rectal wall, and bladder wall for
case 2 are shown in figures 5d-5f.  For this
prostate patient example, there was more separation between the OARs
and the target, and this resulted in a greater potential for
improvement using the MOEA.  The target distributions nearly overlap,
with the clinical plan's coldest 1\% of voxels below 75.5Gy versus
75.4Gy for the MOEA plan, and the clinical plan's hottest 1\% of
voxels above 79.1 Gy versus 79.3 Gy for the MOEA plan.  Dose levels
for both the bladder and especially the rectum are much lower for the
MOEA plan over the entire DVH.   

\vspace{-12pt}
\subsection{Estimation of the Pareto front}
\label{sec:paretofront}

\vspace{-12pt}
The same two cases were used to investigate the ability of the MOEA to
approximate the true Pareto front.  For case 1, a long multiobjective
optimization 
generated 9118 individuals in 12.6 hours without
domination advantage  in order to approximate the true Pareto
front.  Results were compared to 25 shorter optimizations that each
generated 200 individuals using domination advantage. 
Figure 6a plots the results from one of the 25 short runs performed using
domination advantage against the long optimization without using it.
A similar comparison using case 2 was also performed using both a long
(5000 individuals) run and 25 short runs (200 individuals) each.
Results from one of the short runs is plotted with the long run
results in Figure 6b. 

For case 1, each individual in the long
run dominated only $D_C$ = 0.20\% $\pm$ 0.05\% of the
population generated using the short optimization.  For 
case 2, each individual in the true Pareto front
dominated an average of only $D_C$ = 0.14\% $\pm$ 0.04\% of the
population generated using the short optimization.  This demonstrates
how quickly the algorithm with domination advantage approaches the
Pareto front.  The smaller populations were diverse and
covered a wide range of the Pareto front, but certain regions
characterized by diminishing returns were missed when using the
domination advantage feature. 

In order to see if the small performance improvement using domination
advantage in short runs (Section 3.3) translated to longer runs, the
long optimization runs described in the previous paragraph were
repeated with and without domination advantage.  
For case 1, each individual in the long optimization using domination
advantage dominated only $D_C$ = 0.009\% of the population generated
using the long optimization without domination advantage with $D_C$ =
0.004\% in reverse.  For case 2 each
individual produced using domination
advantage dominated only 
$D_C$ = 0.005\% of the population generated using the long
optimization without domination advantage, and the comparison was
$D_C$ = 0.028\% in 
reverse.  Results suggest that for very long runs that closely
approximate the Pareto front, domination advantage
may slow convergence slightly, but differences are very small.  Note
that without the single fitness function domination advantage feature
described in section 2.1.2, the 
MOEA would never get within $\sim\epsilon$ of the 
Pareto front using domination advantage, because all new individuals
only slightly better than the existing population would have been
eliminated.

\vspace{-12pt}
\section{Discussion}

\vspace{-12pt}
We have reported on the development of a hierarchical multiobjective
optimization algorithm that couples a stochastic evolutionary
algorithm with a fast deterministic one.  The multiobjective nature of
the algorithm reflects our view that IMRT planning requires exploring
possible tradeoffs between competing clinical goals.  The decision
making strategy that must be coupled with the the multiobjective
optimization has a profound influence on the structure of the
algorithm.  The current standard clinical decision making strategy
supported by commercial inverse planning systems takes an 
interactive approach in which a plan is generated, evaluated by a
human user, and a new search is made if necessary.  Unfortunately,
these systems do not provide any method for performing an efficient 
and effective search of all possible tradeoffs between objectives.
While the result is usually better than achievable with 3D conformal
techniques, the decision maker is not aware of the large number of
other feasible solutions that may be better.  

Another multiobjective method is the {\it a posteriori} approach in
which a set of Pareto  optimal plans are found first, and then this
set is evaluated by a decision maker to select one.  This is the
nature of our MOEA and is a fairly popular approach for 
multiobjective problems in general \cite{Coello02,Collette03}. It
has the advantage that the user does not have to determine relative
importance amongst objectives before achievable tradeoffs are known.
However, this comes at the price of time since there is little
guidance in the search procedure.  We have used several heuristic methods
to focus the search. In addition, the domination advantage niche effect is
designed to limit the final number of plans to the number that our
decision process can handle.  Bortfeld, Craft and colleagues have
devised an alternative approach in which  only convex objectives are
used, the Pareto front is systematically mapped and calculation times
are reduced by interpolating
solutions\cite{Craft06,Craft07,Hong08,Monz08,Thieke07}.  Specially designed
software allows the decision maker to steer between solutions on the
Pareto front and observe the trade-offs of the different regions.
Another systematic search approach was taken by Lahanas and colleagues
\cite{Lahanas03,Lahanas03b} in which attention was focused on the
method  of completely mapping the Pareto front with a simple decision
making approach used {\it a posteriori}.  An early paper by Yu which
focused on brachytherapy and radiosurgical planning presented an
excellent discussion of multiobjective optimization in radiation
therapy \cite{Yu97b}.  He devised a plan ranking system that
required some {\it a priori} importance input but used a
simulated-annealing method to introduce some fuzziness and to broaden
the search space.   He also included a "satisficing" condition that
functioned similarly to our protocol objectives.  

An {\it a priori} approach to multiobjective decision making in multiobjective
optimization is the third alternative.  Prior methods have the advantages
of being among the fastest and providing a single solution.  Jee {\it et al}
\cite{Jee07} describe a method using the technique of lexigraphic
ordering in which prior ranking of objectives leads to a sequence in
which the first objective is satisfied and then turned into a
constraint as the second objective is optimized and so on for the
remainder.  This is somewhat similar to our shifting the use of our
protocol objectives between the second and third stages, but we make
no preference between the objectives and all are satisfied
simultaneously before transforming them to constraints. 

Our use of an evolutionary algorithm was determined by its extensive
use in multiobjective optimization in 
other fields \cite{Coello02} and because this type of algorithm
can handle a much wider range of  objective types than can
deterministic algorithms.  As with other aspects of our approach, the
trade-off hinges on the speed of convergence.  We used a hierarchical
approach, coupled with the multiple stages, in order to mitigate this
issue, and our results demonstrate that the performance is acceptable
in a clinical setting.  Wu and Zhu used a genetic algorithm for 3D
conformal optimization to optimize the weighting factors \cite{Wu01}.
Although they did not explicitly use Pareto optimality, their choice
of a genetic algorithm was based on a recognition of the
multiobjective nature of the problem and a desire to be flexible with
respect to single objectives.  Evolutionary algorithms have found
their greatest use for the solution of explicitly non-convex
objectives, such as the problem of number and orientation of the radiation
beams \cite{Schreibmann04a, Lahanas03b,Hou03b,  Li04} and the integration of
leaf sequencing and intensity optimization \cite{Li03, Cotrutz03} .
Many other objectives in IMRT planning are 
also non-convex such as dose-volume objectives.  For certain situations, it has been
found that becoming trapped in local minima may be unlikely
\cite{Llacer03} or mathematical techniques are able to be  applied to
linearize non-convex objectives or approximate them with a convex
surrogate \cite{Craft07, Romeijn03}.

A critical component of any multiobjective optimization is the means
by which two objects with multiple characteristics are compared or
rated.  A common example is the comparison of a dose-volume histogram
from two different plans.  When the DVHs do not cross, one can
determine which is clearly better; when they do cross, "better"
depends on the decision variables.
This is the essence of Pareto optimality.  In our IMRT optimization,
we removed plans that were dominated within some epsilon since those
plans were clinically similar or worse in every 
aspect when compared to other plans.  When comparing algorithms or
optimization methods, a similar multiobjective approach can be taken.
In our evaluation of different versions of our 
algorithm, we found it more useful to base our definition of
"better" on a statistical comparison using Pareto dominance as the
defining criterion.  A flexible multiobjective algorithm as described
in this paper is the ideal means for making such comparisons which are
more convincing than alternatives that are commonly used.

\vspace{-12pt}
\section{Conclusion}

\vspace{-12pt}
A multiobjective evolutionary algorithm was developed to find a 
diverse set of Pareto optimal and clinically acceptable IMRT plans.
The novel selection method,
flexible population size, and optimized mutation rates improved
performance.  The population diversity and size was effectively
controlled using a niche effect, "domination advantage".  The protocol objectives and multiple stages used in reproduction and selection guided
the optimization to those regions of plan space that met clinical
criteria.  Key features of the
algorithm are the ability to use any functional form of individual
objectives and the ability to tailor the number and diversity of the
output to the decision making environment.

\vspace{-12pt}
\section{Acknowledgements}

\vspace{-12pt}
Thanks to Stephanie Banerian for the computer and technical
support.  This work was supported by NIH 1-R01-CA112505.

\newpage
%\bibliography{/users/markp/papers/bayes}

\begin{thebibliography}{10}


\bibitem{Meyer04}
J~Meyer, M~H Phillips, P~S Cho, I~Kalet, and J~N Doctor.
\newblock Application of influence diagrams to prostate intensity-modulated
  radiation therapy plan selection.
\newblock {\em Phys Med Biol}, 49:1637--1653, 2004.

\bibitem{Smith09}
W~P Smith, J~Doctor, J~Meyer, I~J Kalet, and M~H Phillips.
\newblock A decision aid for {IMRT} plan selection in prostate cancer based on
  a prognostic {B}ayesian network and a {M}arkov model.
\newblock {\em Artificial Intelligence in Medicine}, 46:119--130, 2009.

\bibitem{Ezzell96}
G~A Ezzell.
\newblock Genetic and geometric optimization of three-dimensional radiation
  therapy treatment planning.
\newblock {\em Med Phys}, 23:293--305, 1996.

\bibitem{Wu01}
X~Wu and Y~Zhu.
\newblock An optimization method for importance factors and beam weights based
  on genetic algorithms for radiotherapy treatment planning.
\newblock {\em Phys Med Biol}, 46:1085--99, 2001.

\bibitem{Coello02}
C~A~C Coello, D~A {Van Veldhuizen}, and G~B Lamont.
\newblock {\em Evolutionary {A}lgorithms for {S}olving {M}ulti-objective
  {P}roblems}.
\newblock Kluwer Academic/Plenum Publishers, New York, 2002.

\bibitem{Osyczka02}
A~Osyczka.
\newblock {\em Evolutionary Algorithms for Single and Multicriteria Design
  Optimization}.
\newblock Physica-Verlag, Heidelberg, Germany, 2002.

\bibitem{Schreibmann04a}
E~Schreibmann, M~Lahanas, L~Xing, and D~Baltas.
\newblock Multiobjective evolutionary optimization of the number of beams,
  their orientations and weights for intensity modulated radiation therapy.
\newblock {\em Phys Med Biol}, 49:747--770, 2004.

\bibitem{Niemierko97}
A~Niemierko.
\newblock Reporting and analyzing dose distributions: A concept of equivalent
  uniform dose.
\newblock {\em Med Phys}, 24:103--110, 1997.

\bibitem{Niemierko99}
A~Niemierko.
\newblock A generalized concept of equivalent uniform dose ({EUD}).
\newblock {\em Med Phys}, 26:1100, 1999.

\bibitem{Luxton08}
G~Luxton, P~J Keall, and C~R King.
\newblock A new formula for normal tissue complication probability ({NTCP}) as
  a function of equivalent uniform dose ({EUD}).
\newblock {\em Phys Med Biol}, 53, 2008.

\bibitem{Breedveld06}
S~Breedveld, P~R~M Storchi, M~Keijzer, and B~J~M Heijmen.
\newblock Fast, multiple optimization of quadratic dose objective functions in
  imrt.
\newblock {\em Physics in Medicine and Biology}, 51, 2006.

\bibitem{Voglis04}
1st International Conference: From Scientific Computing to Computational
  Engineering.
\newblock {\em BOXCQP: An algorithm for bound constrained convex quadratic
  problems}, 2004.

\bibitem{Phillips08}
M~Phillips, M~Kim, and A~Ghate.
\newblock Multiobjective optimization for {IMRT} using genetic algorithm.
\newblock {\em Med Phys}, 35:2753, 2008.

\bibitem{Kalet96}
I~J Kalet, J~P Jacky, M~M Austin-Seymour, and J~M Unger.
\newblock Prism: A new approach to radiotherapy planning software.
\newblock {\em Int J Radiat Oncol Biol Phys}, 36:451--461, 1996.

\bibitem{Kalet97}
I~J Kalet, J~M Unger, J~P Jacky, and M~H Phillips.
\newblock Prism system capabilities and user interface specification, version
  1.2.
\newblock Technical Report 97-12-02, Radiation Oncology Department, University
  of Washington, Seattle, Washington, 1997.

\bibitem{Kalet97a}
I~J Kalet, J~M Unger, J~P Jacky, and M~H Phillips.
\newblock Experience programming radiotherapy applications in {Common Lisp}.
\newblock In D~D Leavitt and G~Starkschall, editors, {\em {XII} International
  Conference on the Use of Computers in Radiation Therapy}, 1997.

\bibitem{Phillips99}
M~H Phillips, K~M Singer, and A~R Hounsell.
\newblock A macropencil beam model: Clinical implementation for conformal and
  intensity modulated radiation therapy.
\newblock {\em Phys Med Biol}, 44:1067--1088, 1999.

\bibitem{Collette03}
Y~Collette and P~Siarry.
\newblock {\em Multiobjective optimization: principles and case studies}.
\newblock Springer Verlag, New York, 2003.

\bibitem{Craft06}
D~L Craft, T~F Halabi, H~A Shih, and T~R Bortfeld.
\newblock Approximating convex pareto surfaces in multiobjective radiotherapy
  planning.
\newblock {\em Medical Physics}, 33, 2006.

\bibitem{Craft07}
D~Craft, T~Halabi, H~A Shih, and T~Bortfeld.
\newblock An approach for practical multiobjective imrt treatment planning.
\newblock {\em Int J Radiat Oncol Biol Phys}, 69, 2007.

\bibitem{Hong08}
T~S Hong, D~L Craft, F~Carlsson, and T~R Bortfeld.
\newblock Multicriteria optimization in intensity-modulated radiation therapy
  treatment planning for locally advanced cancer of the pancreatic head.
\newblock {\em Int J Radiat Oncol Biol Phys}, 72:1208--1214, 2008.

\bibitem{Monz08}
M.~Monz, KH~K{\"u}fer, TR~Bortfeld, and C.~Thieke.
\newblock {Pareto navigation}.
\newblock {\em Physics in medicine and biology}, 53:985--998, 2008.

\bibitem{Thieke07}
C~Thieke, K~H K{\"u}fer, M~Monz, A~Scherrer, F~Alonso, U~Oelfke, P~E Huber,
  J~Debus, and T~Bortfeld.
\newblock A new concept for interactive radiotherapy planning with
  multicriteria optimization: first clinical evaluation.
\newblock {\em Radiother Oncol}, 85:292--298, 2007.

\bibitem{Lahanas03}
M~Lahanas, E~Schreibmann, and D~Baltas.
\newblock Multiobjective inverse planning for intensity modulated radiotherapy
  with constraint-free gradient-based optimization algorithms.
\newblock {\em Phys Med Biol}, 48:2843--2871, 2003.

\bibitem{Lahanas03b}
M~Lahanas, E~Schreibmann, and D~Baltas.
\newblock Intensity modulated beam radiation therapy dose optimization with
  multiobjective evolutionary algorithms.
\newblock In C~M Fonseca, editor, {\em Proc. 2nd Int. Conf. Evolutionary
  Multi-Criterion Optimization, EMO 2003 (Faro, Portugal, 8�11 April 2003)},
  pages 439--47, 2003.

\bibitem{Yu97b}
Y~Yu.
\newblock Multiobjective decision theory for computational optimization in
  radiation therapy.
\newblock {\em Med Phys}, 24(9):1445--54, 1997.

\bibitem{Jee07}
K~W Jee, D~L Mc{S}han, and B~Fraass.
\newblock Lexicographic ordering: intuitive multicriteria optimization for
  imrt.
\newblock {\em Physics in Medicine and Biology}, 52, 2007.

\bibitem{Hou03b}
Q~Hou, J~Wang, Y~Chen, and J~M Galvin.
\newblock Beam orientation optimization for {IMRT} by a hybrid method of the
  genetic algorithm and the simulated dynamics.
\newblock {\em Med Phys}, 30:2360--7, 2003.

\bibitem{Li04}
Y~Li, J~Yao, and D~Yao.
\newblock Automatic beam angle selection in imrt planning using genetic
  algorithm.
\newblock {\em Phys Med Biol}, 49:1915--1932, 2008.

\bibitem{Li03}
Y~Li, J~Yao, and D~Yao.
\newblock Genetic algorithm based deliverable segments optimization for static
  intensity-modulated radiotherapy.
\newblock {\em Phys Med Biol}, 48:3353--3374, 2003.

\bibitem{Cotrutz03}
C~Cotrutz and L~Xing.
\newblock Segment based dose optimization using a genetic algorithm.
\newblock {\em Phys Med Biol}, 48:2987--2998, 2003.

\bibitem{Llacer03}
J~Llacer, J~O Deasy, T~R Bortfeld, T~D Solberg, and C~Promberger.
\newblock Absence of multiple local minima effects in intensity modulated
  optimization with dose-volume constraints.
\newblock {\em Phys Med Biol}, 48:183--210, 2003.

\bibitem{Romeijn03}
H~Romeijn, R~Ahuja, J~Dempsey, A~Kumar, and J~Li.
\newblock A novel linear programming approach to fluence map optimization for
  intensity modulated radiation therapy treatment planning.
\newblock {\em Phys Med Biol}, 48:3521--3542, 2003.

\end{thebibliography}
%\bibliographystyle{unsrt}

%--------------------------- FIGURES -----------------------------------

\begin{figure}
\centering
\subfloat[]{\includegraphics[width=3.1in]{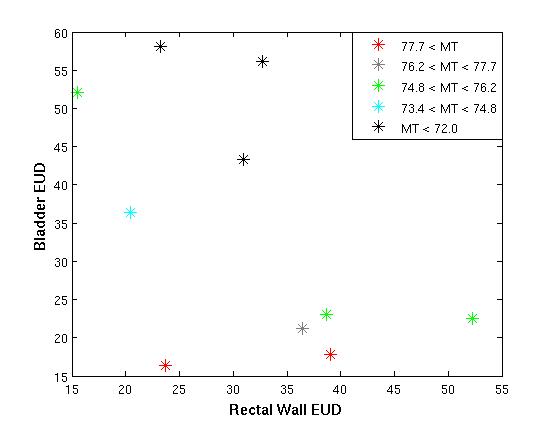}} \\
\subfloat[]{\includegraphics[width=3.1in]{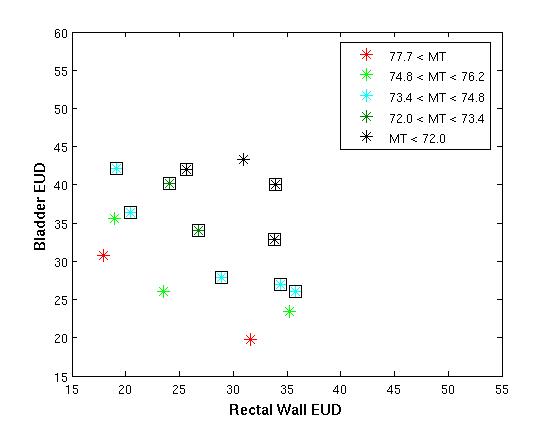}} \\
\subfloat[]{\includegraphics[width=3.1in]{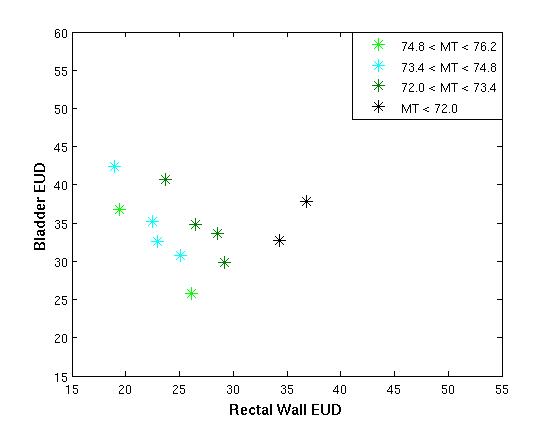}} \\
\caption{Plots of the objective values of the plans at the end of (a)
  stage 1, (b) stage 2, and (c) stage 3.  The colors represent
  different values for the target objective, MT. In (b), the square
  icons represent plans that met the clinical protocol objectives. }
\label{fig:Phases}
\end{figure}

\begin{figure}
\centering
\includegraphics[width=5in]{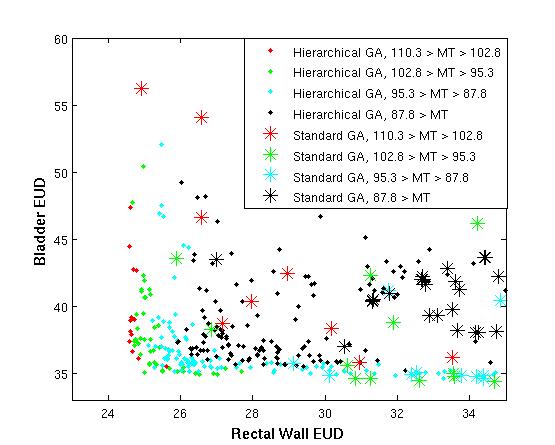}
\caption{Plot comparing the objectives of the final populations of
  plans resulting from the hierarchical approach with that of a
  standard genetic algorithm.  The standard genetic algorithm is shown
  as asterisks, and the hierarchical approach is shown as solid circles.
  The color reflects the target maximum dose (MT), the target
  variance objective.  Lower values in all
  objectives are more optimal.}
\label{fig:GA-vs-hierarchy}
\end{figure}

\begin{figure}
\centering
\includegraphics[width=5in]{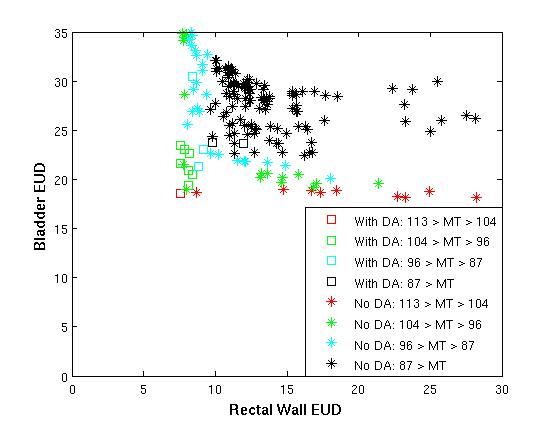}
\caption{Plot demonstrating the effect domination advantage has on a
  population of plans.  The plans resulting from the hierarchical
  approach without using 
  domination advantage are shown as asterisks, and those using domination
  advantage are shown as open squares.  Maximum target dose (MT), the target
  variance objective, is displayed in color.  Lower values in all
  objectives are more optimal.} 
\label{fig:dom-adv}
\end{figure}

\begin{figure}
\centering
\includegraphics[width=5in]{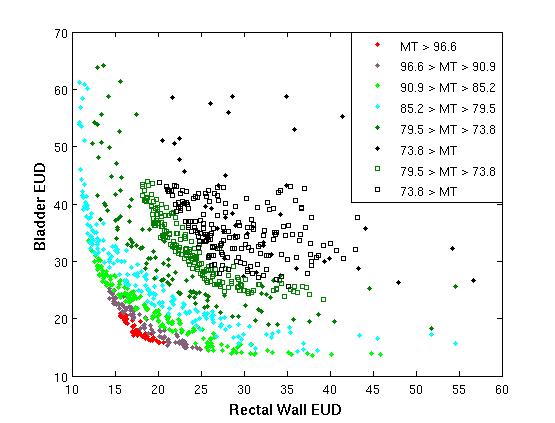}
\caption{Plot comparing multiobjective optimization with (WP) and without (NP) using protocol objectives.  Bladder EUD and
  Rectum EUD are plotted for the sum of 25 full 
  optimizations with and without using protocol objectives to restrict
  the search space.  Maximum target dose (MT), the target variance
  objective, is plotted in color.  The solid circles show the full Pareto
  front, and the open squares display the optimizations using protocol
  objectives.}
\label{fig:protocol-1-obj}
\end{figure}

\begin{figure}
\centering
\subfloat[]{\includegraphics[height=2.5in]{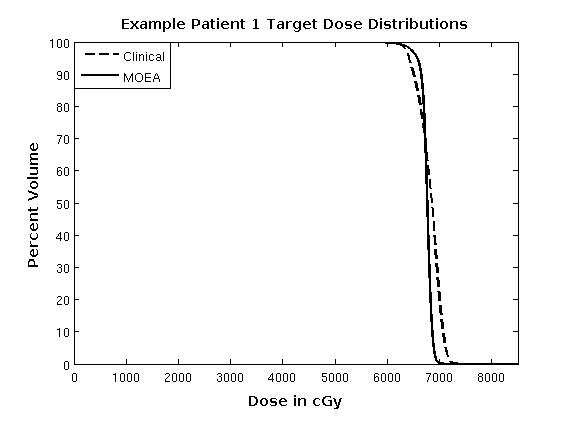}} 
\subfloat[]{\includegraphics[height=2.5in]{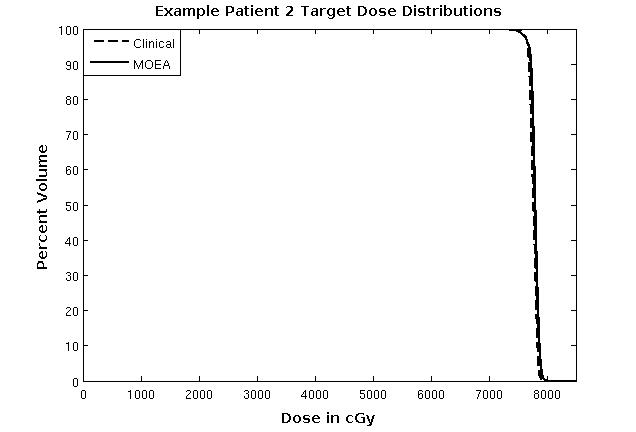}} \\
\subfloat[]{\includegraphics[height=2.5in]{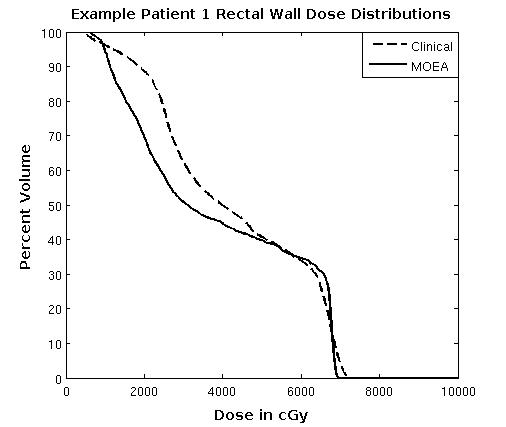}} 
\subfloat[]{\includegraphics[height=2.5in]{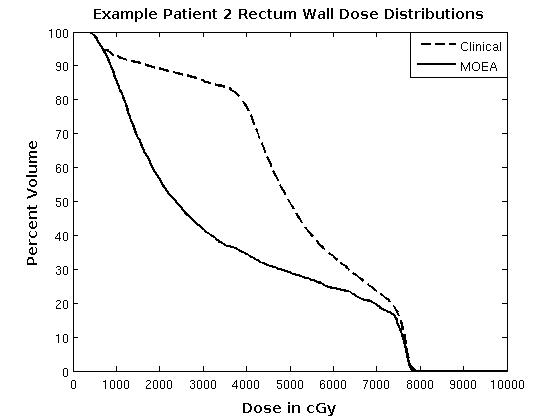}} \\
\subfloat[]{\includegraphics[height=2.5in]{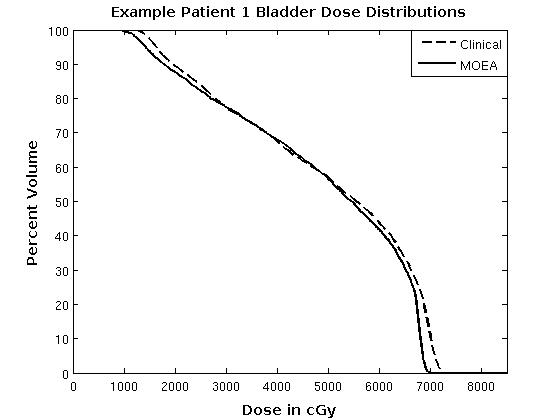}} 
\subfloat[]{\includegraphics[height=2.5in]{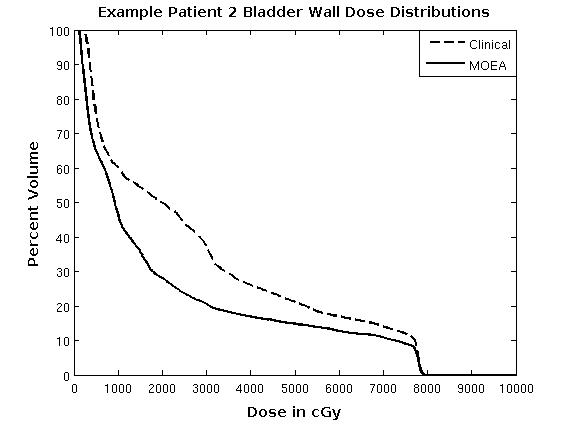}} \\
\caption{DVH comparisons for bladder, bladder wall, rectal wall, and
  PTV are made between plans selected from the population produced by
  the MOEA and plans produced using the current clinical inverse
  planning method.} 
\label{fig:protocol-obj}
\end{figure}

\begin{figure}
\centering
\subfloat[]{\includegraphics[width=3.1in]{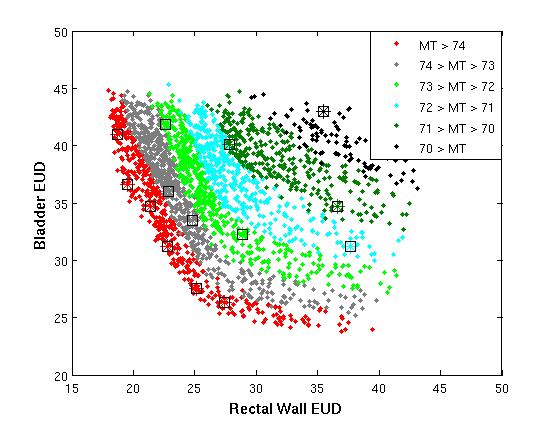}} 
\subfloat[]{\includegraphics[width=3.1in]{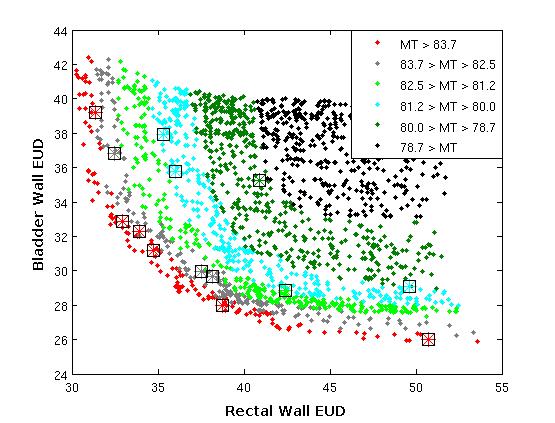}} \\
\caption{Plots for short multiobjective optimizations mapped onto
  an approximation of the Pareto front.  Rectum
  and bladder EUD values are plotted with the target 
  variance objective (MT) shown in color.  The plans from the long
  optimization runs are represented as solid circles that make up most of the
  graphs.  The plans resulting from the quick optimization are
  shown with asterisks using the same colorscale and surrounded by a
  black square for easier viewing.} 
\label{fig:pareto-front}
\end{figure}

\end{document}